\documentclass[preprints,communication,accept,moreauthors,pdftex]{Definitions/mdpi}

\def\D{\mathrm{D}}

\firstpage{1} 
\pubvolume{1}
\issuenum{1}
\articlenumber{0}
\pubyear{2022}
\copyrightyear{2022}
\externaleditor{Academic Editor: %MDPI: please add if available.
}
\datereceived{} 
\dateaccepted{} 
\datepublished{} 
\hreflink{https://doi.org/} % If needed use \linebreak

\Title{Can Dark Energy Emerge from a Varying $G$ and \mbox{Spacetime Geometry?}}

\TitleCitation{Can Dark Energy Emerge from a Varying $G$ and Spacetime Geometry?}

\Author{Ekim %MDPI: Please carefully check the accuracy of names and affiliations.
T.  %MDPI: This author is different from Redmine. Please confirm.
Han{\i}meli $^{1,}$*\orcidA{} , Isaac Tutusaus $^{2,3,4,5}$, Brahim Lamine $^{5}$ and Alain Blanchard $^{5}$}

\AuthorNames{Ekim T. Han{\i}meli, Isaac Tutusaus, Brahim Lamine and Alain Blanchard}

\AuthorCitation{Han{\i}meli, E. T.; Tutusaus, I.; Lamine, B., Blanchard, A.}
\address{%
$^{1}$ \quad Zentrum f{\"u}r Angewandte Raumfahrt und Mikrogravitation (ZARM), Universit{\"a}t Bremen, Am Fallturm 2, 28359 Bremen, Germany\\
$^{2}$ \quad Universit\'e de Gen\`eve, D\'epartement de Physique Th\'eorique and Centre for Astroparticle Physics, 24 quai Ernest-Ansermet, CH-1211 Gen\`eve 4, Switzerland; isaac.tutusauslleixa@unige.ch\\ 
$^{3}$ \quad Institute of Space Sciences (ICE, CSIC), Campus UAB, Carrer de Can Magrans, s/n%Please confirm whether this could be removed.
, 08193 Barcelona, Spain\\
$^{4}$ \quad Institut d’Estudis Espacials de Catalunya (IEEC), Carrer Gran Capit\`a 2-4, 08034 Barcelona, Spain\\
$^{5}$ \quad Institut de Recherche en Astrophysique et Plan\'etologie (IRAP), Universit\'e de Toulouse, CNRS, UPS, CNES, \mbox{14 Av. Edouard} Belin, F-31400 Toulouse, France; brahim.lamine@irap.omp.eu (B.L.); alain.blanchard@irap.omp.eu (A.B.)
\\

}

\corres{Corresponding author: ekim.hanimeli@zarm.uni-bremen.de}

\abstract{The accelerated expansion of the universe implies the existence of an energy contribution known as dark energy. Associated with the cosmological constant in the standard model of cosmology, the nature of this dark energy is still unknown. We
will discuss an alternative gravity model in which this dark energy contribution emerges naturally, as a result of allowing for a time-dependence on the gravitational constant, $G$, in Einstein's field equations. With this modification, Bianchi's identities require an additional tensor field to be introduced so that the usual conservation equation for matter and radiation is satisfied. The equation of state of this tensor field is obtained using additional constraints, coming from the assumption that this tensor field represents the space-time response to the variation of $G$. We
will also present the predictions of this model for the late-universe data, and show that the energy contribution of this new tensor is able to explain the accelerated expansion of the universe without the addition of a cosmological constant. Unlike many other alternative gravities with varying gravitational strength, the predicted $G$ evolution is also consistent with local observations and therefore this model does not require screening. We 
will finish by discussing possible other implications this approach might have for cosmology and some future prospects.}


\begin{document}

%%%%%%%%%%%%%%%%%%%%%%%%%%%%%%%%%%%%%%%%%%
%\setcounter{section}{-1} %% Remove this when starting to work on the template.

%\begin{paracol}

\section{Introduction}

Modern cosmology has increasingly become a precision science during
the last \mbox{30 years,} allowing the determination of many properties of the universe to a great degree of accuracy~\cite{blanchard,d.weinberg}. As it has happened many times in the history of physics, this new abundance of data has also created various tensions within the standard framework~\cite{challenges, bull}. 

One of the major puzzles cosmologists are facing today is the so called cosmological constant problem~\cite{s.weinberg,carroll1992}. A helpful way of looking at this problem is as the standard theory's inability to satisfactorily explain the nature of the ``dark energy'' that drives the observed accelerated expansion of the universe.
In standard cosmology, dark energy is represented by a cosmological constant term, $\Lambda$, within Einstein's equations. A constant added this way to the equations behaves exactly as the vacuum energy, which was, historically, first noticed by Lema\^itre \cite{lemaitre}. 
General relativity has no natural reason to produce this term, so that it has to be added ad hoc. Fortunately, quantum field theory expects that there should indeed be a vacuum energy. The problem is that, even only taking into account the energy scales for which we know quantum field theory works, without referring to anything unknown, the calculated value of vacuum energy density is many orders of magnitudes higher than what is measured with the expansion of the universe.

This large vacuum energy does not create problems in the quantum field theory, because quantum theory allows one to ignore the zero-point-energy, and only energy differences become important. For general relativity, on the other hand, all energy, including the zero-point-energy, should gravitate. The question of how the quantum vacuum should gravitate in itself highlights a deep incompatibility between quantum theory and general relativity. 

It is possible to imagine that a theory of quantum gravity will have some sort of a mechanism allowing vacuum energy density to be set to a value that matches the cosmological observations.%MDPI: Footnote is not permitted in this journal, so we have moved it into Notes before references, please confirm the whole text.
\endnote{It is possible that this may be achieved through a more careful treatment of quantum vacuum within the current semi-classical paradigm as proposed by \cite{Unruh2017,Unruh2018}. This approach, however, is subject to ongoing debate \cite{bengochea2020}.} However, even assuming this would be the case, the question ``Why should the vacuum energy density be this far too small, but otherwise non-remarkable value instead of zero'' still remains. To this, broadly, three types of response are possible:
 
  1. There is nothing necessarily problematic about the value of the cosmological constant. Just like the gravitational constant, or fine structure constant, energy density of the vacuum is this value measured by the expansion rate of the universe. Quantum theory should simply incorporate this constant. 
  
  2. Dark energy is not really the vacuum energy, but some other particle field that looks like a cosmological constant at the first approximation. Cosmologists should then try to create models for this substance and understand its properties.
  
  3. Gravity does not work exactly as general relativity envisions, and there should be a modification in such a way to explain away the accelerated expansion. In this view, dark energy is the artefact of a gravitational mechanism.

In the following, we will discuss the implications of an often overlooked possibility within the third option. We will investigate a model where the gravitational constant, $G$, changes with cosmological time, and we will argue that, in this picture, dark energy that drives the cosmological expansion emerges necessarily, as a response of the spacetime geometry. It can be noted that various modified gravity theories (Brans--Dicke being the prime example \citep{bransdicke}) also involve an effective variation of the gravitational constant. In comparison to these, the $G(t)$ %Please check throughout to make sure that all signs/symbols/values, etc. are kept in same format.
model we will discuss has the advantage of being more conservative, in the sense that it makes no additional assumptions about the nature of gravity beyond general relativity. In fact, as there is no fundamental reason that $G$ should be a constant, this model can be said to lift that assumption. This way, instead of offering a new theory of gravity, we study a phenomenological extension to general relativity. Furthermore, we will show that the varying $G$ model is not only consistent with the late-universe data, but also the expected fluctuation in $G$ is small enough to be consistent with independent measurements of the gravitational constant such that no screening of gravitational interactions is needed.

In Section \ref{sec2} we will begin by demonstrating how an additional term is required by the geometry of 
spacetime if $G$ is not a constant. We will follow this up by outlining the derivation of the cosmological equations including the energy contribution from this additional term. In Section \ref{sec3}, we will discuss the cosmological data used in this research, paying particular attention to how supernovae should be affected by a change in the gravitational strength, and present the results of our analysis. We will then conclude with a discussion of the results, and what other impacts this model might have for cosmology. The discussion and results presented in the following are 
based on Ref.~\cite{prd} by the present authors, and we refer the reader to this reference for further details.

%%%%%%%%%%%%%%%%%%%%%%%%%%%%%%%%%%%%%%%%%%
\section{Varying  \textbf{\emph{G}} in Einstein's Equations}\label{sec2}

Our starting point is Einstein's equations with $G$ being a time-dependent variable, namely: 

\begin{equation}
G^{\mu \nu}= 8\pi G(t, \vec{x}) T^{\mu \nu}\,,
\end{equation}
with $G^{\mu\nu}$ and $T^{\mu\nu}$ being the Einstein and stress-energy tensors, respectively. According to the cosmological principle in the following, we take $G(t, \vec{x})=G(t)$. 

However, with equations in this form, Bianchi's identities imply that the energy density of the normal constituents, such as matter or radiation, become dependent on the gravitational constant: 
\vspace{-10pt}

\begin{equation}
\D_\mu G^{\mu\nu} = 0 \quad\Rightarrow\quad\D_\mu T^{\mu\nu} = -\frac{T^{\mu \nu} \partial_\mu G(t)}{G(t)} \not = 0  \quad\Rightarrow\quad \rho_{\text{matter}} \propto G^{-1} a^{-3}\,,
\end{equation}
where $\rho$ stands for the energy density and $a$ represents the scale factor.

One simple way that preserves both the conservation of energy and Bianchi's identities is if there is an additional dynamical term in the equations. In general, this will be a rank-2 tensor which satisfies 

\begin{equation}
\D_{\mu}S^{\mu \nu}=-T^{\mu \nu}\partial_{\mu}G\,.
\label{eq1}
\end{equation}

We see, then, if $G$ is a time-dependent variable in Einstein's equations, $\D_\mu G^{\mu\nu} = 0$ and $\D_\mu T^{\mu\nu} = 0$ together necessitate a modification of the form

\begin{equation}
G^{\mu \nu}= 8\pi G(t)T^{\mu \nu}+ 8\pi S^{\mu\nu}\,,
\end{equation}
where the factor $8\pi$ is added for convenience.

With the usual cosmological assumptions, isotropy and homogeneity, this tensor\textbf{$S^{\mu\nu}$} can be written in terms of only two scalar functions, $\Phi(t)$ and $\Psi(t)$,

\begin{equation}
S^{\mu \nu}=(\Phi(t) +\Psi(t))u^\mu u^\nu + \Psi(t) g^{\mu \nu}\,,
\end{equation}
where 
$g^{\mu\nu}$ represents the metric. The 4-velocity, $u^{\mu}$, is taken as the Hubble flow, as we take $S^{\mu \nu}$ to have no bulk motion with respect to space.

Note that the $\Phi(t)$ component works exactly as an energy contribution, which we need to determine in order to compare this model with cosmological observations. In fact, Equation~(\ref{eq1}) is a general expression that should also apply to a variety of dark energy fluid models, which would come with a specific $\Phi$. In our case, we determine this from the argument that the tensor $S^{\mu \nu}$ does not represent a fluid, but instead represents the spacetime response to the varying $G$. 

We can use Equation~(\ref{eq1}) above to find an expression between $\Phi(t)$ and $\Psi(t)$, but this is not easy to solve for $\Phi$, even if $\Psi(t)$ is known. However, we can put the equations in a more useful form by defining an auxiliary function $\xi(t)$ such that, $\dot{\xi}/\xi=3H(1+\Phi/\Psi)$, with $H$ being the Hubble rate. Then, in terms of this function we can integrate \mbox{Equation~(\ref{eq1}) as}

\begin{equation}
\Phi(t) \xi(t)=\lim_{\varepsilon \to 0} \bigg(\Phi(\varepsilon) \xi(\varepsilon)  - \int_\varepsilon^t \dot{G} \rho \xi dt \bigg)\,,
\label{phi}
\end{equation}

\begin{equation}
\xi(t)= \lim_{\varepsilon \to 0} \Bigg[ \xi(\varepsilon) \frac{a(t)^{3\big(1+\frac{\Phi(t)}{\Psi(t)}\big)}}
{a(\varepsilon)^{3\big(1+\frac{\Phi(\varepsilon)}{\Psi(\varepsilon)}\big)}}
\exp\bigg(-3\int_{\varepsilon}^t  \frac{d(\Phi/\Psi)}{dt}\ln(a) dt \bigg) \Bigg]\,.
\end{equation}

These do not look much more promising, but they allow us further simplifications. First, notice that the scale factor, $a$, dependence appears exclusively through the function $\xi$. Second, the first equation above includes two constant terms coming from the lower limit of the integral. We want to argue that the energy contribution of $S^{\mu \nu}$, the term we added to preserve the geometrical consistency, is exclusively caused by varying $G$. Therefore, we should insist that $\Phi$ (and, by the virtue of Equation~(\ref{eq1}), $\Psi$) 
is zero if $G$ is a constant. This implies that all the constants in Equation~(\ref{phi}) should be zero. Furthermore, $\xi$ should be a constant, otherwise $\Phi$ would have an additional scale factor dependence, beyond what is induced by $G$ and $\rho$.

With these simplifications, we can obtain an expression for $\Phi$ in terms only of $G$ and other cosmological parameters. For $G$, we use an ansatz in the form of a power series centred around the scale factor $a=1$, of which we use only first few terms in the analysis, 

\begin{equation}
G(a)=G_0\Bigg(1 + \sum_{n=1}^{\infty} b_n (1-a)^n \Bigg)\,.
\end{equation}

Using the flat Friedmann--Lema\^itre--Robertson--Walker (FLRW) metric, we obtain the cosmological equations in the usual manner:

\begin{equation}
H^2 = \frac{8 \pi G\rho}{3}+\frac{8 \pi \Phi}{3}, \quad \quad \frac{\ddot{a}}{a}= -\frac{4 \pi G}{3} (\rho + 3p) -\frac{4 \pi}{3}(\Phi + 3 \Psi)\,,
\end{equation}
with $p$ representing the pressure of the fluid.

%%%%%%%%%%%%%%%%%%%%%%%%%%%%%%%%%%%%%%%%%%
\section{Analysis and Results}\label{sec3}

As our model is concerned with dark energy, we use late-universe probes, namely type-Ia supernovae (SNIa) and baryon acoustic oscillations (BAO). For BAO, we use measurements from 6dFGS~\cite{6dfgs}, SDSS-MGS~\cite{sdss}, BOSS DR12~\cite{dr12}, and eBOSS DR14~\cite{dr14}, Ly-$\alpha$ autocorrelation function~\cite{bautista}, and Ly-$\alpha$-quasar cross-correlation~\cite{dumas} at $z=2.4$. For SNIa, we use the JLA dataset \citep{jla}.

As we do not assume screening, supernova luminosities should be modified as they would be affected by the variation in gravitational strength. A straightforward argument, popular in the cosmology literature, is to include a $G$ dependence of luminosity through Chandrasekhar's mass. By assuming that a higher mass of explosion will lead to a higher nickel content and, therefore, higher luminosity, one has $L \propto M_{Ch} \propto  G^{-1.5}$~\cite{gaztanaga}. On the other hand, the recent literature \cite{wrightli,sakstein} on supernovae physics indicates a relationship on the opposite direction, $L \propto  G^{1.46}$. In our study, we use both approaches, but as the predicted $G$ variation is quite small we do not obtain a notable difference between the two. Therefore we only report the results from the first approach.

Using a $\chi^2$ minimization analysis, we determine the cosmological parameters: matter density, $\Omega_m$, radiation density, $\Omega_r$, and the product of the Hubble constant times the standard ruler, $H_0 r_d$; as well as the parameters of $G$: $b_1$, $b_2$, and $b_3$. Of the latter, only two are independent due to flatness (we take the dependent parameter to be $b_1$). Higher order terms for $G$ do not improve the fit, as they are less relevant for $a$ close to one. The $\chi^2$ values and best-fit values obtained for all parameters %results 
can be found in Equations~(\ref{tab1})--(\ref{tab3}). 
As Equation~(\ref{tab1}) shows, the present varying $G$ model is consistent with cosmological data in a similar way to $\Lambda$CDM:

\begin{equation}
       \chi^2_{\text{Varying G}}/\text{d.o.f.}= 697.73/747 , \quad \quad      \chi^2_{\Lambda CDM}/\text{d.o.f.} = 698.05/749\,.\label{tab1}
\end{equation}

\begin{equation}
       b_1 = 0.07 \pm 0.15 , \quad \quad  b_2=  -0.51 \pm 0.33 ,  \quad \quad  b_3= 0.679 \pm 0.094 \,.\label{tab2}
\end{equation}

\begin{equation}
       \Omega_m = 0.284 \pm 0.017 ,   \Omega_r= (0.0 \pm 7.0) \times 10^{-3},     H_0 r_d = (101.7 \pm 1.3) \times 10^{2}  \text{[km s$^{-1}$]}\,.\label{tab3}
\end{equation}

\begin{figure}[H]
\includegraphics[width=.98\linewidth]{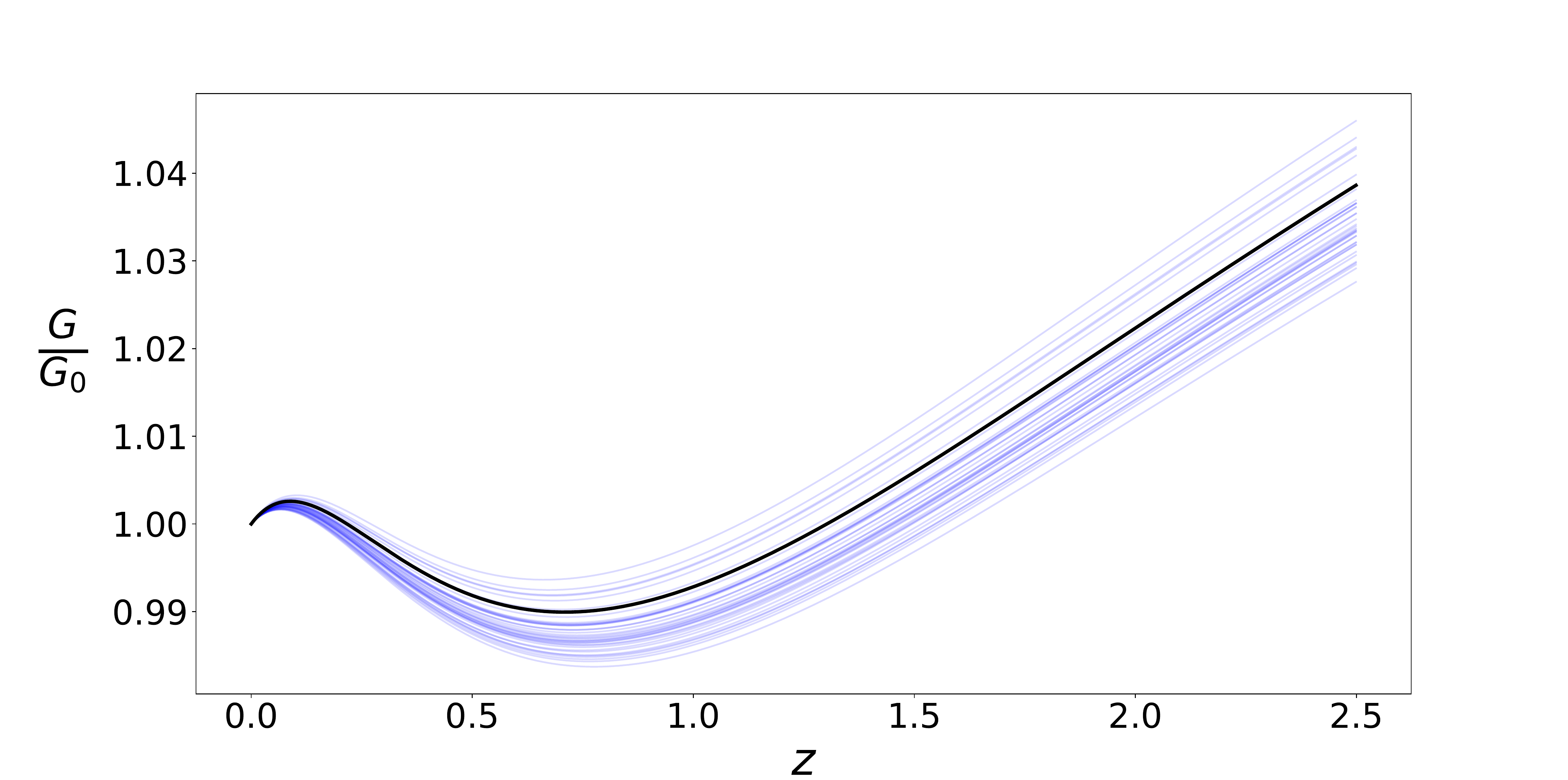}
\caption{Evolution of $G$ vs redshift. The line is drawn using the best fit values for $b_1$, $b_2$, and $b_3$ while the blue lines are some sample error lines with one $\chi^2$ difference ($\Delta \chi^2=1$) from the black line.
\label{fig1}}
\end{figure}

When we look at the $G$ evolution in Figure~\ref{fig1}, we see that the expected variation of $G$ is below 5\% of the standard value in the epochs of interest. It appears that the change of $G$ increases as we go to earlier times, but this is largely due to the function used as an ansatz for $G$. As we include no data from the early-universe, our present results have no predictive value from these times.

An interesting parameter to note is $b_1$, which is equal to $-\frac{\dot{1}}{H_0}\frac{\dot{G}}{G}$ at $a=1$. As independent observations of $G$ measure $\frac{\dot{G}}{G}$ (for instance Lunar Laser Ranging \cite{llr}), this parameter can be used to estimate if the present measurements of $G$ are consistent with our model (alternatively, it is possible to constrain $b_1$ using $\frac{\dot{G}}{G}$ measurements). We see that $b_1$ is consistent with zero, meaning that, under the present model, no small-scale modifications are required to the gravitational interactions for consistency with local observations.

\section{Conclusions and Outlook}\label{sec4}

In the preceding discussion we have considered the possibility that the observed ``dark energy'' is emergent as a result of a variation of $G$, due to the geometry of spacetime in Einstein's equations. This allows the interpretation that dark energy can be formulated as the response of the spacetime to the varying $G$. 
Comparing our model with cosmological observations from the late universe, we have found that the considered model is consistent with the data in a similar way to the standard cosmology. This indicates that this model can adequately explain the accelerated expansion of the universe without a cosmological constant, avoiding one part of the cosmological constant problem. 

Furthermore, we have found that the necessary variation of $G$ is relatively small, which means that the small-scale interactions do not need to be ignored (or screened), in contrast to many other modified gravity approaches. As this model is also very conservative in terms of assumptions, it can be a good basis for studying possible changes of gravitational strength on astrophysical scales.

Of course, we do not claim that this model can compete the standard model in terms of its completeness, and this investigation can be extended in various ways. From a cosmological perspective, an immediately interesting question is whether a varying $G$ model is compatible with the early-universe data. For this, the most important indicator is cosmic microwave background (CMB) constraints on the variation of $G$. However, this analysis is not straightforward and requires a treatment of the perturbations. Along with CMB, constraints from structure growth and big bang nucleosynthesis can also be included in the analysis to have a fuller understanding of the constraints on $G$ beyond the late-universe we have focused on. Such an analysis may also have the effect of 
decreasing other tensions within modern cosmology, an example being the $H_0$ value. A cursory look at the cosmological equations suggests that, at the background level, a smaller $G$ at CMB epoch can accommodate a larger $H_0$ value without changing the expansion history. While it is possible that such a change in $G$ will not be compatible with the observations, a further analysis of the present model may bring insights into the $H_0$ problem as well.

%%%%%%%%%%%%%%%%%%%%%%%%%%%%%%%%%%%%%%%%%%
\vspace{6pt} 

\authorcontributions{E.T.H. led the analysis and preparation of the manuscript. I.T., B.L., and A.B. contributed to the analysis and preparation of the manuscript.
 %For research articles with several authors, a short paragraph specifying their individual contributions must be provided. The following statements should be used ``Conceptualization, X.X. and Y.Y.; methodology, X.X.; software, X.X.; validation, X.X., Y.Y. and Z.Z.; formal analysis, X.X.; investigation, X.X.; resources, X.X.; data curation, X.X.; writing---original draft preparation, X.X.; writing---review and editing, X.X.; visualization, X.X.; supervision, X.X.; project administration, X.X.; funding acquisition, Y.Y. All authors have read and agreed to the published version of the manuscript.'', please turn to the  \href{http://img.mdpi.org/data/contributor-role-instruction.pdf}{CRediT taxonomy} for the term explanation. Authorship must be limited to those who have contributed substantially to the work~reported..
}

\funding{ This research received no external funding %Please add: ``This research received no external funding'' or ``This research was funded by NAME OF FUNDER grant number XXX.'' and  and ``The APC was funded by XXX''. Check carefully that the details given are accurate and use the standard spelling of funding agency names at \url{https://search.crossref.org/funding}, any errors may affect your future funding..
}

\institutionalreview{Not applicable. %In this section, you should add the Institutional Review Board Statement and approval number, if relevant to your study. You might choose to exclude this statement if the study did not require ethical approval. Please note that the Editorial Office might ask you for further information. Please add “The study was conducted in accordance with the Declaration of Helsinki, and approved by the Institutional Review Board (or Ethics Committee) of NAME OF INSTITUTE (protocol code XXX and date of approval).” for studies involving humans. OR “The animal study protocol was approved by the Institutional Review Board (or Ethics Committee) of NAME OF INSTITUTE (protocol code XXX and date of approval).” for studies involving animals. OR “Ethical review and approval were waived for this study due to REASON (please provide a detailed justification).” OR “Not applicable” for studies not involving humans or animals..
}

\informedconsent{Not applicable. %Any research article describing a study involving humans should contain this statement. Please add ``Informed consent was obtained from all subjects involved in the study.'' OR ``Patient consent was waived due to REASON (please provide a detailed justification).'' OR ``Not applicable'' for studies not involving humans. You might also choose to exclude this statement if the study did not involve humans. Written informed consent for publication must be obtained from participating patients who can be identified (including by the patients themselves). Please state ``Written informed consent has been obtained from the patient(s) to publish this paper'' if applicable..
}

\dataavailability{All the data relevant to this article can be found through the cited works.  %In this section, please provide details regarding where data supporting reported results can be found, including links to publicly archived datasets analyzed or generated during the study. Please refer to suggested Data Availability Statements in section ``MDPI Research Data Policies'' at \url{https://www.mdpi.com/ethics}. If the study did not report any data, you might add ``Not applicable'' here..
}

\conflictsofinterest{The authors declare no conflict of interest.  %Declare conflicts of interest or state ``The authors declare no conflict of interest.'' Authors must identify and declare any personal circumstances or interest that may be perceived as inappropriately influencing the representation or interpretation of reported research results. Any role of the funders in the design of the study; in the collection, analyses or interpretation of data; in the writing of the manuscript, or in the decision to publish the results must be declared in this section. If there is no role, please state ``The funders had no role in the design of the study; in the collection, analyses, or interpretation of data; in the writing of the manuscript, or in the decision to publish the~results''..
}

\begin{adjustwidth}{-\extralength}{0cm}
%\centering %% If there is a figure in wide page, please release command \centering
\printendnotes[custom]
\reftitle{References}

% Please provide either the correct journal abbreviation (e.g., according to the “List of Title Word Abbreviations” http://www.issn.org/services/online-services/access-to-the-ltwa/) or the full name of the journal.


\end{adjustwidth}

\end{document}